%% file: proconda.tex
\documentclass{llncs}
\usepackage{epsfig,endnotes}
\usepackage[compatibility=false]{caption} 
\usepackage[]{hyperref}
\usepackage{amsmath}
\usepackage{amssymb}
\usepackage{subfig}
\usepackage[title]{appendix}
\input{style}

\newcommand{\proconda}{\textsc{ProConDa}\xspace}

\newcommand{\dsg}{Data Source Graph\xspace}
\newcommand{\phaseone}{Control Data Identification}
\newcommand{\phasetwo}{Write Instruction Identification}
\newcommand{\phasethree}{Control Data Write Instruction Origin (WIO) Checking}

\graphicspath{{./figures/}}

\usepackage{etoolbox}
\makeatletter
\patchcmd{\hyper@makecurrent}{%
    \ifx\Hy@param\Hy@chapterstring
        \let\Hy@param\Hy@chapapp
    \fi
}{%
    \iftoggle{inappendix}{
        \@checkappendixparam{chapter}%
        \@checkappendixparam{section}%
        \@checkappendixparam{subsection}%
        \@checkappendixparam{subsubsection}%
        \@checkappendixparam{paragraph}%
        \@checkappendixparam{subparagraph}%
    }{}%
}{}{\errmessage{failed to patch}}

\newcommand*{\@checkappendixparam}[1]{%
    \def\@checkappendixparamtmp{#1}%
    \ifx\Hy@param\@checkappendixparamtmp
        \let\Hy@param\Hy@appendixstring
    \fi
}
\makeatletter

\newtoggle{inappendix}
\togglefalse{inappendix}

\apptocmd{\appendix}{\toggletrue{inappendix}}{}{\errmessage{failed to patch}}
\apptocmd{\subappendices}{\toggletrue{inappendix}}{}{\errmessage{failed to patch}}

\begin{document}


\date{}

\title{\Large Proconda\\ \textbf{Pro}tected \textbf{Con}trol \textbf{Da}ta}

\author{\vspace*{1cm}
{Marie-Therese Walter}
\and
{David Pfaff}
 \and
 {Stefan N\"urnberger}
\and
 {Michael Backes}
} 


\institute{CISPA Helmholtz Center for Information Security\\
Saarland Informatics Campus
}

\maketitle

\subsection*{Abstract}
Memory corruption vulnerabilities often enable attackers to take control of a target system by overwriting control-flow relevant data (such as return addresses and function pointers), which are potentially stored in close proximity of related, typically user-controlled data on the stack. 

In this paper, we propose \proconda, a general approach for protecting control-flow relevant data on the stack

\proconda leverages hardware features to enforce a strict separation between control-flow relevant and regular data of programs written in non-memory-safe languages such as C.
Contrary to related approaches, \proconda does not rely on information hiding 
and is therefore not susceptible to several recent attacks specifically targeting information hiding as a foundation for memory isolation. 

We show that \proconda enforcement is compatible with existing software by applying a software-based prototype to industry benchmarks on an ARM CPU running Linux.

\input{sections/introduction}
\input{sections/attackermodel}
\input{sections/design}
\input{sections/implementation}
\input{sections/evaluation}

\input{sections/relatedwork}
\input{sections/conclusion}

\bibliographystyle{acm}
\bibliography{proconda}

\begin{subappendices}
	\renewcommand{\thesection}{\Alph{section}}%
	\input{sections/controldataidentification}
	\input{sections/armassembly}
	\input{sections/programcoverage}
\end{subappendices}


\end{document}

%% file: style.tex
\usepackage{listings} 
\usepackage{caption} 
\usepackage{color} 
\usepackage{enumerate} 
\usepackage{xspace}

\definecolor{darkgray}{gray}{0.5}
\definecolor{lightgray}{gray}{0.9}
\definecolor{darkred}{rgb}{0.7,0,0}
\definecolor{darkgreen}{rgb}{0,0.7,0}
\definecolor{darkblue}{rgb}{0,0,0.7}

\newcommand{\code}[1]{\texttt{#1}}

\DeclareCaptionFont{white}{\color{white}}
\DeclareCaptionFormat{listing}{\colorbox{darkgray}{\parbox{
	\linewidth-2\fboxsep}{#1#2#3}}}
			
\lstdefinelanguage
[x86]{Assembler}     
[x86masm]{Assembler} 
{alsoletter={\%},
morekeywords=[1]{%
CDQE,CQO,CMPSQ,CMPXCHG16B,JRCXZ,LODSQ,MOVSXD, %
POPFQ,PUSHFQ,SCASQ,STOSQ,IRETQ,RDTSCP,SWAPGS, %
movl, movzbl, addl, subl, pushl, popl},
morekeywords=[2]{%
\%si,\%al,%
\%eax,\%ebx,\%ecx,\%edx,\%esi,\%edi,\%esp,\%ebp,\%eip,%
\%rax,\%rdx,\%rcx,\%rbx,\%rsi,\%rdi,\%rsp,\%rbp,\%rip,%
\%r8,\%r8d,\%r8w,\%r8b,\%r9,\%r9d,\%r9w,\%r9b, %
}} 

\lstdefinelanguage[ARM]{Assembler}%
   {morekeywords=[1]{MOV,ADD.W,B.W},
  alsoletter={.,0,1,2,3,4,5,6,7,8,9},%
    alsodigit={?},%
    sensitive=false,%
    morestring=[b]",%
    morecomment=[s]{/*}{*/},%
    morecomment=[l]@,%
    morecomment=[l]//,%
   }[keywords,comments,strings]

%% file: sections/introduction.tex
\section{Introduction}
\label{sec:introduction}
In the last several years, a seemingly unstoppable amount of attacks have surfaced that exploit buffer overflows, dangling pointers and other related memory corruption vulnerabilities.
They all have in common that the attacker manages to divert the control flow of a vulnerable program, thereby enabling them to either skip code (e.g. password checks), run a function that was not supposed to be executed (e.g. copy file) or even run an arbitrary chain of instructions using Return-Oriented Programming (ROP)~\cite{sh07:innocentflesh} techniques.

While the majority of these vulnerabilities could be avoided by using memory-managed languages (e.g. Java), many features of unmanaged languages have become deeply ingrained in existing software and cannot be adapted due to performance and portability concerns. Similarly, enhancing existing software with memory safety features often comes at a significant cost~\cite{nazhma10:CETS}. 
Because of these obstacles, many early defensive mechanisms against memory corruption vulnerabilities focused on providing light-weight security guarantees with only small impacts on performance. 

\emph{Data Execution Prevention (DEP)} was one of the first popular, low-overhead protection mechanisms. It successfully prevents any type of code-injection attacks by preventing an attacker-controlled call stack from carrying executable code. However, the stack can still hold control-flow relevant data such as function pointers and return addresses. Hence, the attacker can still gain control of the execution path and can selectively direct control flow to appropriate code locations marked as executable. Such techniques are commonly referred to as \emph{code re-use attacks}.

The most fundamental approach for addressing code re-use attacks is \emph{Control Flow Integrity (CFI)}~\cite{abbuer05:cfi,abbuer09:cfi,dahapa15:hafix}. At each control flow branch, CFI checks whether the actual control flow target is still identical to the originally intended control flow.
Unfortunately, this approach can only detect the attack at a point after the control flow data has already been corrupted and the program counter is about to jump to an unintended location. At this point, the memory, and therefore the state of the program, have long been tampered with and it is not possible for the program to recover into a known-good state. 

The only way to stop all memory corruption exploits is to enforce complete memory safety~\cite{szpawe13:sok} as done by memory-managed languages like Java. For unmanaged languages like C, a multitude of defenses have been proposed that either try to enforce spatial or temporal memory safety. While approaches focused on spatial memory safety mostly work by restricting pointer or object bounds, temporal memory safety approaches usually use additional allocation information to check if a pointer or object is still valid~\cite{szpawe13:sok}. 

Like CFI, however, these approaches also 
do not defend against the initial corruption of the control flow data. Therefore, program recovery is impossible. 

\subsection{Our Approach}

In this paper, we present \emph{\proconda}, a method for protecting control data that identifies and prevents control flow exploits one stage earlier than CFI, namely when the control-flow relevant data (\emph{control data}) is manipulated on the stack. \proconda consists of a general framework that uses a fine-grained form of access control to ensure the integrity of control data when the data is written.

The key idea of \proconda is to detect and prevent unintended writes to control data based on \emph{who} (which instruction) attempts to write control data. \proconda utilizes the memory management unit (MMU) of the processor to trip on write attempts to control data on the stack and then detect \emph{which} instruction attempted to write control data. The code location of this write instruction---the \emph{Write Instruction Origin (WIO)}---enables us to distinguish between instructions that are actually allowed to write control data and unwanted side effects of another memory access. The negative performance impact of checking each write attempt is avoided by augmenting legitimate write instructions with code that announces the imminent changes. This ensures that all other write instructions, which did not announce their intent to write to control data on the stack, will have their access to these data denied. 
The actual control flow change, i.e., reading control data and jumping to the respective address, does not need to be protected since \proconda ensures that it has not been tampered with. Therefore, \proconda renders additional CFI checks obsolete.

\subsection{Contribution}
In detail, this paper makes the following contributions:
\begin{itemize}
	\item We present a method to systematically and programmatically identify legitimate control data writes. 
	\item We present an effective method to check the write instruction origin (\emph{WIO}) by rewriting programs such that they announce upcoming legitimate writes, thereby making it possible to catch un-announced illegal write attempts. 
	\item Identifying illegitimate write attempts rather than their consequences such as unintended control flow changes, allows us to pinpoint the exact location of an error, i.e. the location of the vulnerable code snippet, and hence enables more targeted bug fixing.
	\item As a proof of concept, we provide a prototypical software-only implementation of our approach targeting the ARM architecture for Linux platforms.
\end{itemize}

%% file: sections/attackermodel.tex
\section{Attacker Model}
\label{sec:attackermodel}
 Similar to Code Pointer Integrity~\cite{kuszpa14:cpi}, we assume that an attacker can fully control the process memory but cannot modify the code segment, as it is by default write-protected.
 In contrast to traditional techniques like ASLR, \proconda does not employ a form of information hiding as a defense mechanism but rather arranges control data (i.e. data that will eventually be used as input to control flow instructions) in a separate, write-protected memory region. Since writing to this region is restricted to legitimate write instructions and therefore prohibited for all other instructions, we assume that an attacker also cannot manipulate this special memory region to divert control flow.
 
 Additionally, we assume that an attacker cannot read the value of special registers reserved by the compiler and that the kernel does not save any user-level registers in user accessible memory when memory switches occur. This is congruent with the assumptions from CCFI~\cite{mabibo15:ccfi}. 
 
Like for other CFI-related approaches, the focus of our approach also lies on control-flow hijacking attacks that are mounted by exploiting memory corruption errors in the program. Hence, we aim to prevent the attacker from gaining control over control-flow relevant registers such as the the program counter (\code{pc}) or the link register (\code{lr}), which is used to handle return addresses in ARM.

To achieve this goal, \proconda enforces WIO checks to guarantee the integrity of data used to influence these registers.
\proconda therefore defends against memory corruption attacks, such as buffer overflow exploits, format string exploits, or dangling pointers. As a result, \proconda protects programs against a range of return-oriented programming flavors such as counterfeit object-oriented programming~\cite{scteli15:coop} or control-flow bending~\cite{cabapa15:CFbending}.

%% file: sections/design.tex
\section{Design}
\label{sec:design}

\proconda aims at providing access control for write attempts to control-flow relevant data on the stack by preventing access attempts, which are not reflected in the program semantics. Any attempt to abuse general purpose write instructions to overwrite control data is detected and prevented. 

While it is hard for static analysis to enumerate where control flow might jump to, it is possible to accurately identify the positions in code where control flow is changed and where control data is written. We use this observation to implement access control based on \emph{which} instruction is attempting to overwrite control data. If a write access is attempted by an instruction that normally does not write control data, it is considered to be unintended and is prevented by \proconda in-line checks during run-time.

\proconda uses the code locations of write instructions to distinguish them. An access control check for write attempts can therefore be reduced to finding the write instruction's code location in a list of statically extracted allowed locations. Consequently, manipulations of control data originating from a location not on this list must be a result of write attempts intended for other data. We call the code location of a write instruction the \emph{Write Instruction Origin (WIO)}.

\begin{figure*}
	\centering
	\includegraphics[width=0.9\textwidth]{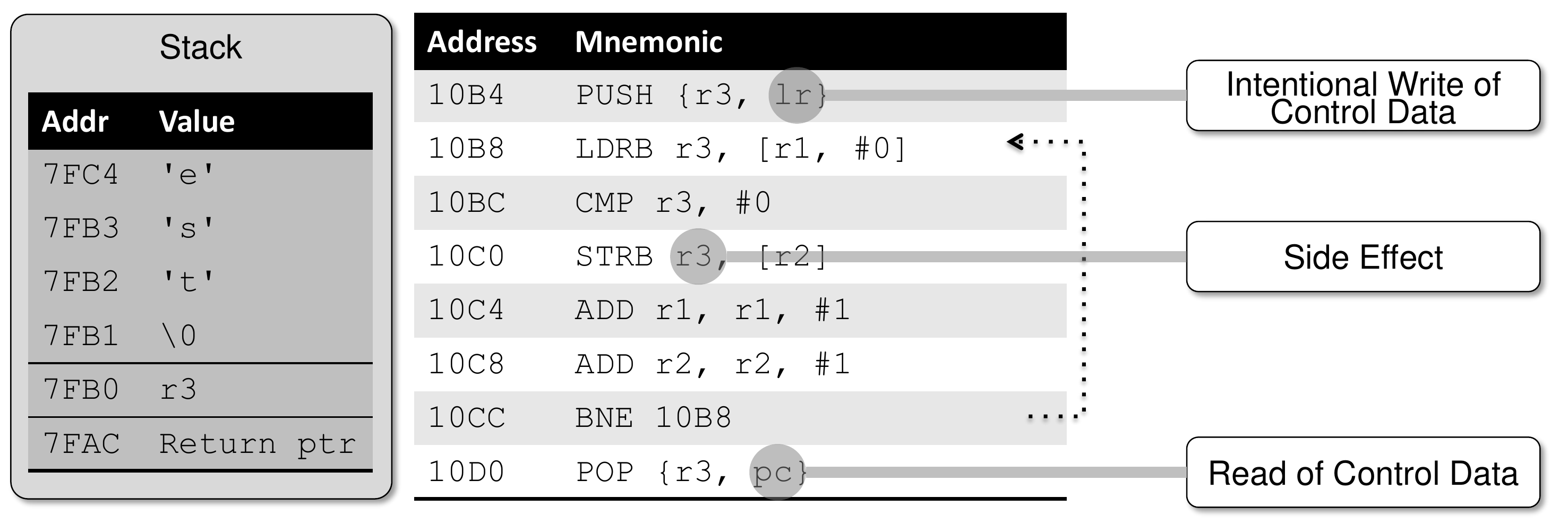}
	\vspace*{0.3cm}
	\caption{Disassembly of function \code{strcpy} and its stack layout.}
	
	\label{fig:assemblyexample}
\end{figure*}

Consider the example snippet of ARM assembly for the function \code{strcpy} depicted in \autoref{fig:assemblyexample}. This function is vulnerable to a buffer overflow because it does not check if the destination pointer \code{r2} exceeds the allocated space of the destination buffer on the stack. In the example in \autoref{fig:assemblyexample}, the destination buffer is four bytes long and, when overflown, will point to the stored value of \code{r3} and eventually to the return pointer at \code{0x7FAC}. 

To better understand this vulnerability, let us analyze the code snippet in reverse, starting at the last instruction: \code{POP \{r3, pc\}}. The instruction \code{POP}s the value stored at address \code{0x7FAC} from the stack into the program counter \code{pc}. Using this address, which is of course only valid in the scope of \code{strcpy}, we can track back which part of the code writes data to that particular address by means of static analysis. In this case, the \code{PUSH} instruction at \code{0x10B4} is the only instruction that is intended to write to \code{0x7FAC}. 
Therefore, any other write attempt to \code{0x7FAC} constitutes an unintended write. \proconda builds upon this fact and identifies such illegitimate write attempts. 

\proconda is designed to operate in three phases. The first phase statically analyses the program to find control-flow relevant data (control data) on the stack, while the second phase uses backwards slicing to identify which instructions are intended to write to that control data. The first two phases only need to be applied once and produce a rewritten version of the program's assembler code with the desired write origin checks embedded in the code. The third phase takes place during the execution of the program. After compiling the program using the newly secured assembler files, \emph{WIO} checks are automatically performed by the program at run-time.

\subsection{Phase 1: \phaseone}
\label{subsec:design1}

In the first phase, \proconda identifies all vulnerable control data on the stack by inspecting the set of instructions that are able to change control flow. To this end, \proconda distinguishes between three types of instructions:

\begin{enumerate}
	\item Instructions that only advance control to the next instruction (fall-through)
	\item Instructions with a hard-coded control-flow destination 
	\item Instructions whose destination is supplied by a register or memory location
\end{enumerate}

The first two types are considered harmless as their control-flow destination is hard-coded in the program and the code segment is write-protected. Type 3, however, is of our interest as the destination could have been tampered with by an attacker. A register or memory location used by such an instruction contains control data. This data directly influences the control flow.
Since control data is not syntactically different from any other data, it can only be identified by inspecting its usage. 

\begin{figure*}
	\centering
	\includegraphics[width=0.9\textwidth]{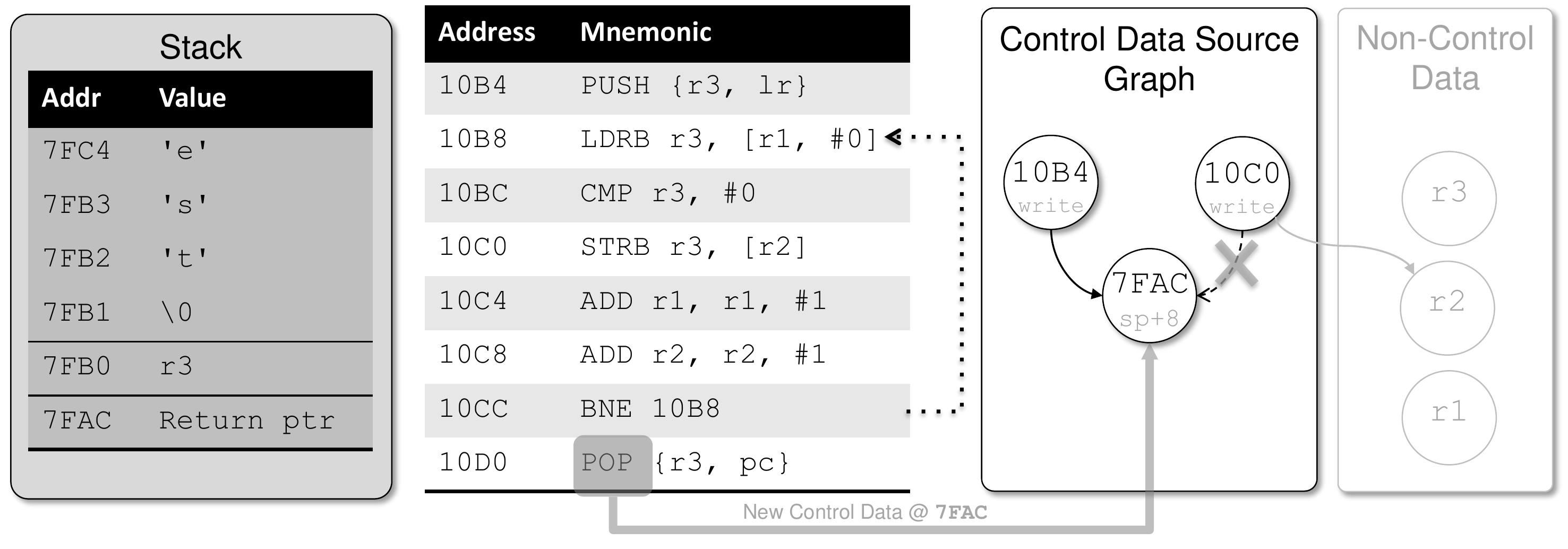}
	\vspace*{0.3cm}
	\caption{Data Source Graph of the function \code{strcpy} shown in \autoref{fig:assemblyexample}.}
	\label{fig:data_source_graph}
\end{figure*}

In the example in \autoref{fig:data_source_graph}, the only instruction that changes control-flow is the \code{POP} instruction stored at \code{0x10D0}. This instruction pops an address from the stack and overwrites the current program counter \code{pc}, thereby changing control flow. Hence, the second value on the stack can be classified as being control data, because it is used as control flow data when this instruction is executed. By marking the logical position as control data, we can use backwards slicing to detect where it was written (phase 2) in order to find the corresponding write instruction and its WIO, the code location of the instruction relative to the entry point of the program. Later in phase 3, this WIO is used to decide whether a write attempt is allowed or not.

Unlike the example here, there are cases where the control-flow destination does not stem from memory directly. An overview of these cases and how they are handled by \proconda can be found in appendix \ref{appendix:cdi}.

\subsection{Phase 2: \phasetwo}
In the second phase of \proconda, write instructions that target the discovered memory locations containing control data are identified using backwards slicing. Since several write instructions may target the same memory location, we represent the discovered relation as a directed graph, the \emph{\dsg}: each write instruction has a directed edge pointing to the control data it manipulates. The direction of the edge corresponds to the write direction. The \dsg for our running example is shown in  \autoref{fig:data_source_graph}. 

In this example, the control data on the stack, which is the return address of the function \code{strcpy}, can only be written by the \code{PUSH} instruction stored at \code{0x10B4}. Consequently, \code{0x10B4}, the address of the write instruction, has an edge pointing to \code{0x7FAC}, the address of control data, in the \dsg.  The write instruction at \code{0x10C0} on the other hand should not manipulate the control data at \code{0x7FAC} and therefore does not have an edge in the \dsg pointing to this data.

In phase 3, the \dsg can be used during execution of the program to decide if a write access is intended or not. If the corresponding WIO has an edge pointing to the affected control data in the \dsg, the write access is intended otherwise it is considered unintended.

\subsection{Phase 3: \phasethree}
During execution of the program, each write access to control data (\code{0x7FAC} in the example above) is monitored (WIO checking), while read access is always allowed. Since the integrity of control data is checked on write attempts to that data, enforcing an additional check on read access to that data is not necessary. 

If the \dsg shows that a write attempt was made by an instruction that does not have an edge pointing to the control data address in question, it triggers a fault. 
In the \dsg depicted in \autoref{fig:data_source_graph}, an unintended write attempt by the \code{STRB} instruction stored at \code{0x10C0} is not reflected as an edge in graph and hence would be caught. The program is then intentionally aborted as the program state would be undefined after the control data has been overwritten.

%% file: sections/implementation.tex
\section{Implementation}
\label{sec:implementation}

To show the feasibility and effectiveness of our approach, we implement a software-based prototype of \proconda for an ARM CPU running a Linux kernel.
Our prototype consists of a tool-chain that first statically analyses the assembler files of a program to identify potentially vulnerable code pointers and their locations before using this information to dynamically identify legitimate write instructions to these pointers using hardware memory watchpoints.
An explanation for choosing assembly as a program representation for our analysis and for deciding on the ARM as the underlying hardware section can be found in~\autoref{appedinx:taa}.

In the following, we describe how our prototype implements the three \proconda phases in detail.

\subsection{\phaseone}
For the first phase, we statically analyze a program's assembler files to identify registers that are used in a control-flow relevant instruction, e.g. registers used in branch instructions as shown in~\autoref{fig:control_data_write}. In this example, we have a branch instruction at location \code{0x10D4} using register \code{r3}. In contrast to static labels that represent fixed code locations, the register can at this point contain an arbitrary address that might have been influenced by user input earlier on. Therefore, we mark the register as containing control data and remember the code location of the branch instruction as its offset from either the entry point of the program, if the instruction happens in the \code{main} function, or from the label of the function that contains the instruction. This way of representing code locations is made possible by the fact that in assembler, function names are represented as unique labels that can be used to jump to specific points in the program. Therefore, \code{<functionLabel> + <offset>} can be used to uniquely identify instructions by their code locations, independent their actual memory address which might differ from run to run due to protection mechanisms like Address Space Layout Randomization (ASLR).

We use the information about the register and the code location of the instruction to identify which \emph{base address} has been loaded into the register before the instruction is executed. In case of a chain of dereferencing instructions, the base address is the last loaded address before the start of the dereferencing chain. In all other cases, the base address is simply the last loaded address. The reason for this differentiation between address types is that our prototype aims only at protecting control data that is normally stored on the stack. 

Once we have identified at which point in the program the base address is loaded into the register, we remember the code location of the corresponding load instruction. By remembering this location instead of the actual base address, we avoid falsely marking memory addresses as control-flow relevant in case protection mechanisms like ASLR are active.

In our example, we have read instructions using register \code{r3} at \code{0x10CC} and at  \code{0x10C8}. Since the read instruction at \code{0x10CC} dereferences register \code{r3} by loading the content of the address contained in \code{r3} into \code{r3} itself, the base address in this example can be found in the instruction at \code{0x10C8} where the content of \code{sp-4} is loaded into \code{r3}. We therefore remember code location \code{0x10C8} as the location of the instruction that loads the base address into register  \code{r3}. We repeat this process for all identified, control-flow relevant read instructions, resulting in a list of code locations that load potentially control-flow relevant data (control data) at runtime.

\subsection{\phasetwo}
\label{subsec:implphasetwo}
\begin{figure}
	\centering
	\includegraphics[width=\linewidth]{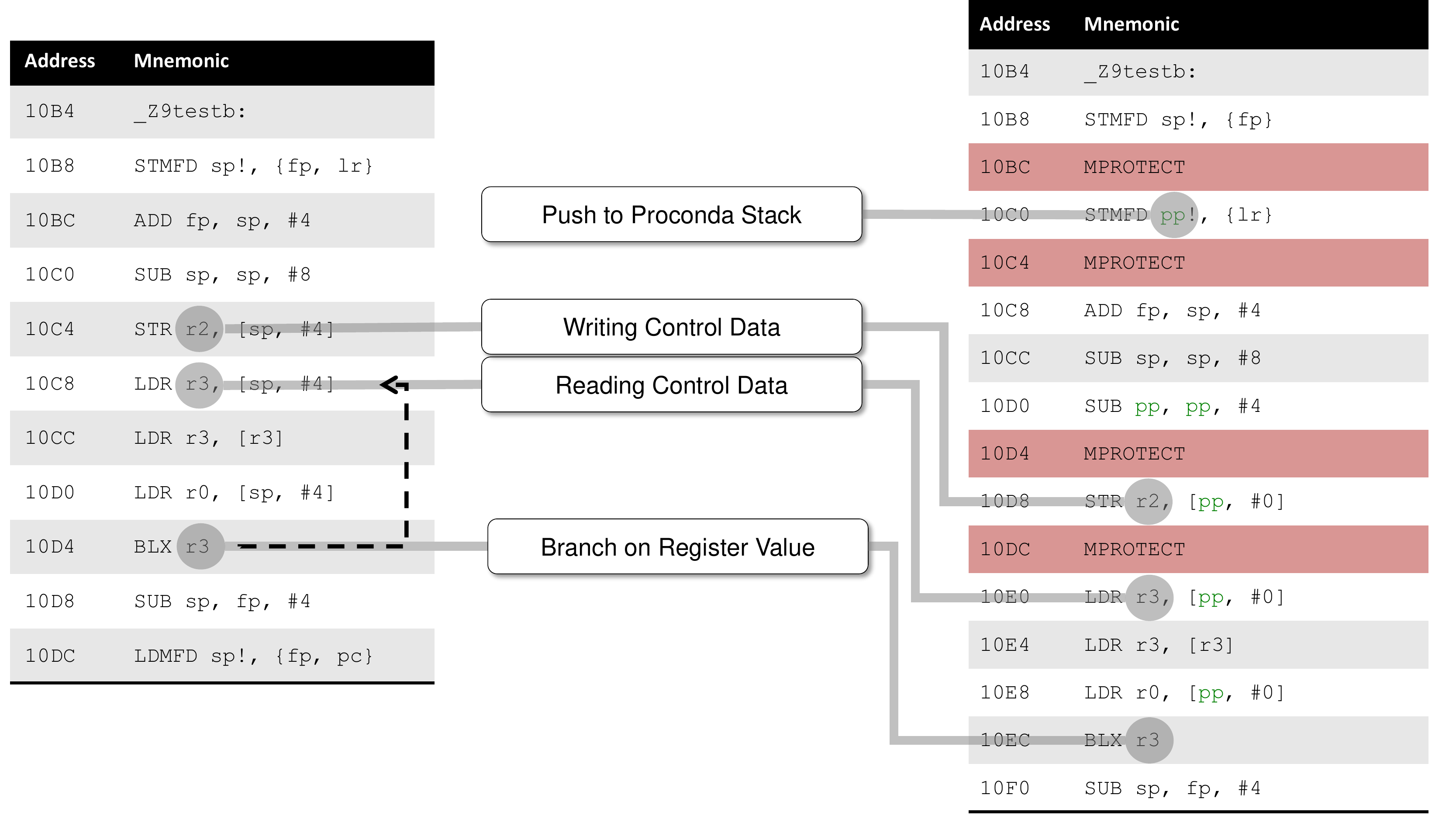}
	\caption{Identification of legitimate write instructions targeting control-flow relevant data and the secured \proconda assembly.}
	\label{fig:control_data_write}
\end{figure}

In the second phase, we generate backwards slices for all control-flow relevant addresses loaded by instructions at the code locations  we identified in phase one. 
To handle inter-procedural jumps and to restrict access to control-flow relevant data to the instructions that actually require this access, we use the GNU Debugger (GDB) to dynamically generate the backwards slices. GDB has the advantage that it disables ASLR by default, i.e. running the program several times will always yield the same control-flow relevant addresses. This lack of randomization allows us to set watchpoints on all control-flow relevant addresses and then run the program to identify the code locations of legitimate write instructions for these addresses. As soon as an instruction changes the content of an address with a watchpoint, GDB will pause the program execution and save the instruction that caused the change along with its code location, resulting in a list of legitimate instructions and their code locations. 
As shown in \autoref{fig:control_data_write}, for our example, this step results in the write instruction at \code{0x10C4} being identified as a legitimate write of control data. An explanation of how \proconda achieves sufficient program coverage in this step can be found in~\autoref{appendix:aopc}.

After identifying legitimate write instructions using GDB, we use this information to rewrite the program's assembler files. First, we move all vulnerable code pointers along with all return pointers to a special \textit{.proconda} section that is by default read-only and stored on a separate memory page. We then add \code{mprotect} calls before and after each legitimate write instruction to temporarily make the \textit{.proconda} section writable just for that instruction. 
Finally, we update all other legitimate references to code pointers contained in read instructions to reflect their move to the \textit{.proconda} section and compile the rewritten assembler files into executable code. The result for our example is shown in \autoref{fig:control_data_write}.

The reason for combining \code{mprotect} calls with a separate, write-protected memory page in our prototype is that, at the time of its implementation, hardware supported memory access control mechanisms like Intel MPK had been announced but were not publicly available yet. Therefore, we used the aforementioned combination to simulate a one-level access control mechanism. Since protection mechanisms like Intel MPK~\cite{intelsys} may offer up to 16 access levels, future \proconda implementations could enforce a more fine-grained and efficient access control mechanism beyond the capabilities of our current prototype.

\subsection{\phasethree}
\label{subsec:implphasethre}
This final phase takes place during execution of the newly secured program: since only pre-identified, legitimate write instructions are allowed to actually access the protected \textit{.proconda} section, each attempt at modifying the data outside of these instructions, e.g. by exploiting a buffer overflow vulnerability will result in a segmentation fault, effectively crashing the program.

Pending on access to hardware features such as Intel MPK, future implementations of \proconda will defer the handling of denied write attempts either to the operating system or to the program itself to avoid these segmentation faults and program crashes.

%% file: sections/evaluation.tex
\section{Evaluation}
\label{sec:evaluation}
In this section we verify that the \proconda mechanism is able to protect programs with memory corruption vulnerabilities by subjecting our prototypical implementation to real-world exploits. Furthermore, we evaluate the performance impact of the general idea of WIO checking as well as the performance impact and the space requirements of its software emulation based on \code{mprotect} calls.

\subsection{Runtime overhead evaluation}
\label{sec:evaluation:perf}

\begin{figure*}
	\subfloat[Software-simulated \proconda%
	\label{fig:softref}]{%
		\includegraphics[width=0.45\textwidth]{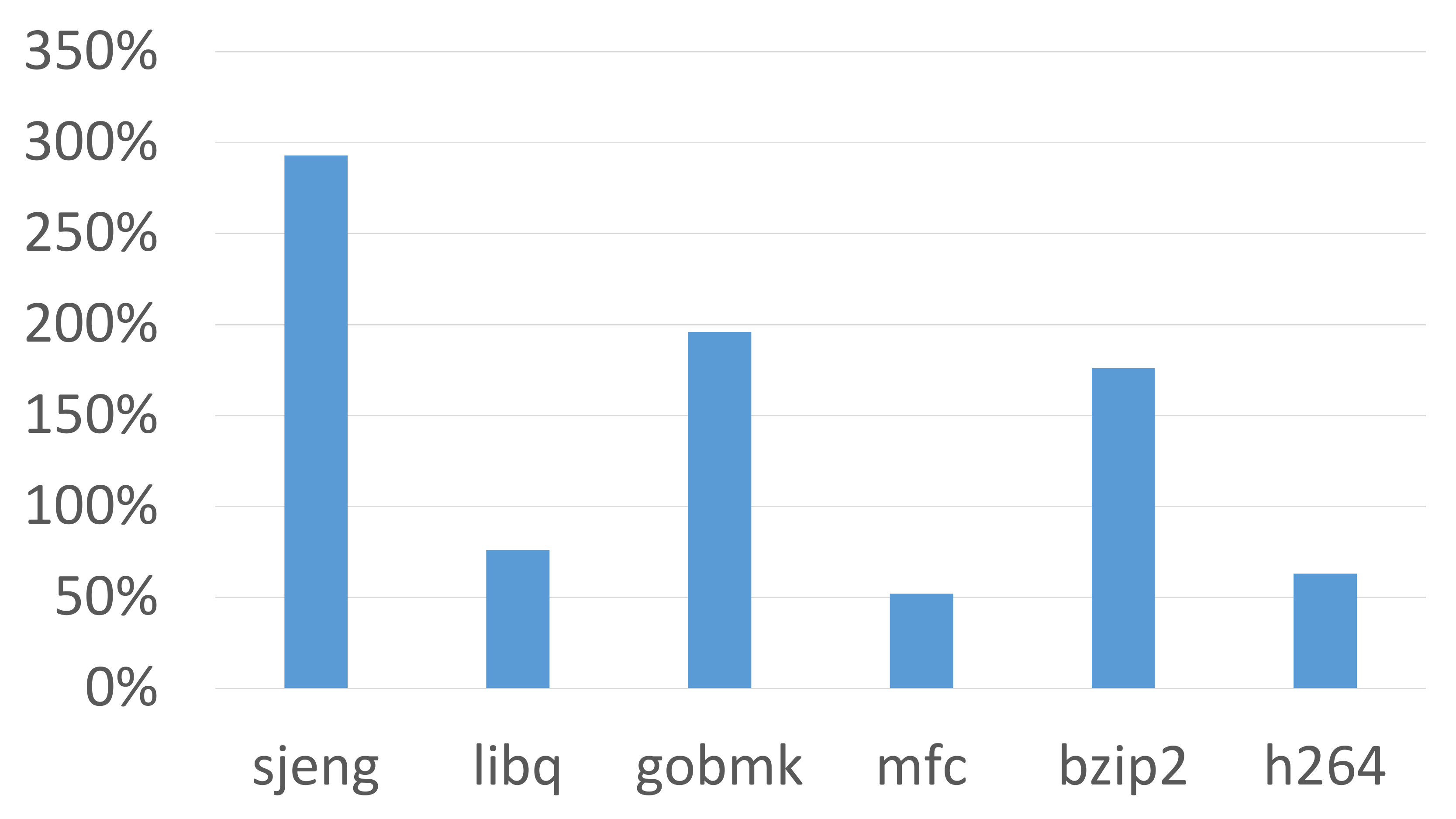}}
	\hfill
	\subfloat[Reference lower bound without \code{mprotect} calls%
	\label{fig:hwref}]{%
		\includegraphics[width=0.45\textwidth]{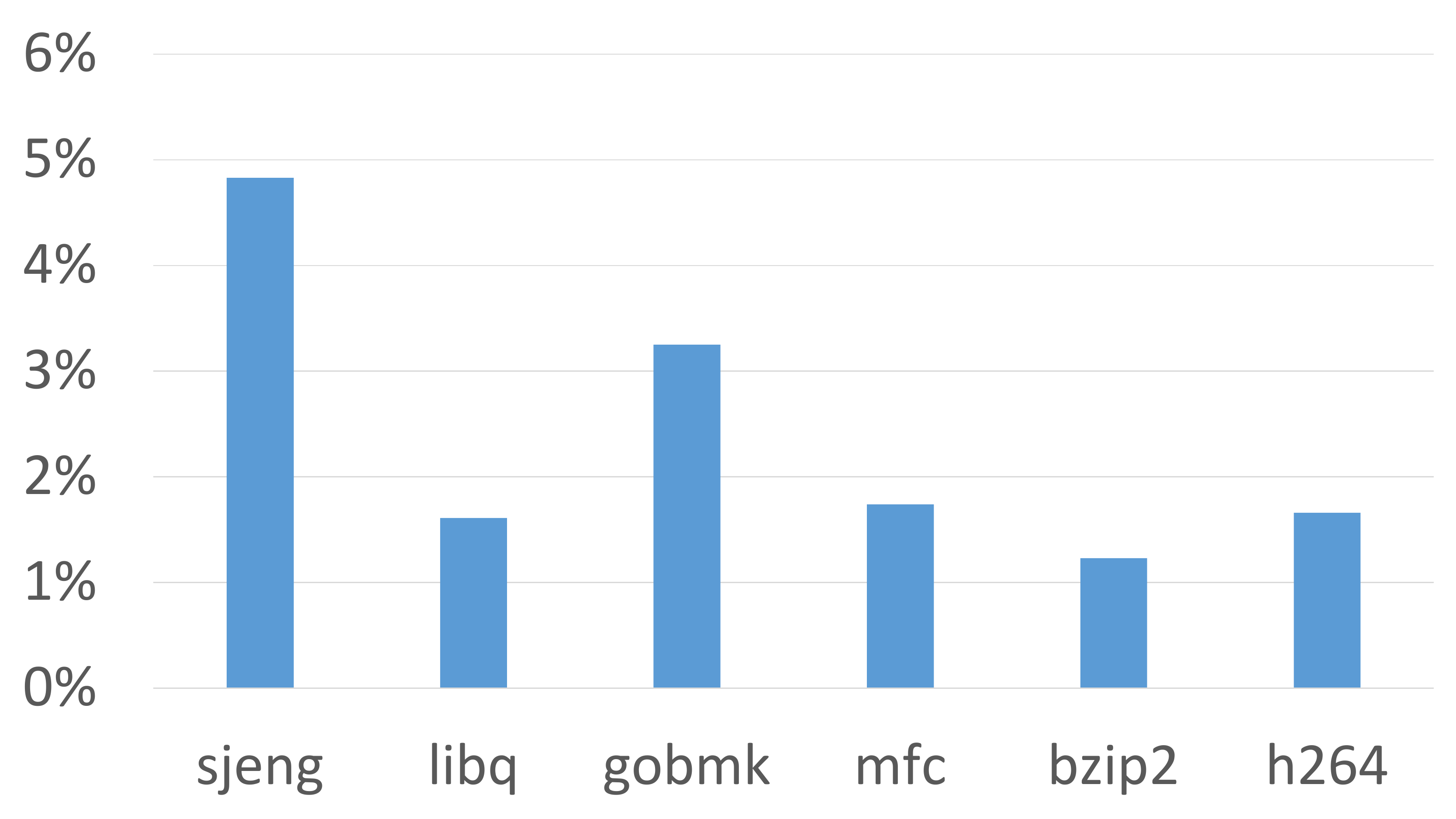}}
	\caption{\proconda run-time overhead using SPEC CPU2006.}
	\label{fig:1}
\end{figure*}

The general idea behind \proconda -- distinguishing between different origins of instructions that write control data -- is versatile and hence can be implemented in different ways. While an optimal solution would rely on hardware features to separate legitimate writes from illegitimate ones, current processors do not support such a feature yet. 
Intel MPX shares \proconda's goal of protecting against memory corruption exploits, it uses bounds checking as its protection mechanism and is therefore of no interest to us.

Since hardware support is not yet available, we utilize existing, software-based tools\footnote{In our case, we use the ability to create separate, read-only memory pages for certain data and to selectively make these pages writable using \code{mprotect} calls.} to emulate and enforce memory access control. This is why in this section, we perform two separate performance benchmarks: (1) an upper bound estimate with a software-only emulation of memory protection mechanisms implemented using \code{mprotect}, and (2) a lower bound estimation of the general WIO idea that replaces the non-yet-existing hardware instructions with \code{NOP}s. The idea behind using \code{NOP}s is that, in contrast to \code{mprotect} calls, \code{NOP}s and hardware instructions do not cause context switches, which are costly in terms of performance. Except for the WIO checking instructions (\code{mprotect} vs. \code{NOP}), both performance benchmarks are set up identically.

We measured the performance overhead of the software simulation implementing \proconda on some of the SPEC CPU2006 integer benchmarks~\cite{sp07:spec}. All experiments where performed on a Raspberry Pi 2 BCM2836 ARM processor chip running a Debian 8 OS with Linux kernel version 4.1.17. For the SPEC benchmark, we selected all programs which were compatible with our ARM environment out-of-the-box and whose output was validated as the correct result. The reported overhead is the average of 5 runs, the measured variance was negligible (standard deviation less than 0.5\%).

The benchmarks were instrumented with WIO checks according to the \proconda principle. 
The relative slow-down in performance gives the total overhead of the \proconda solution. The instrumentation for benchmarks (1) and (2) differ as  (1) is implemented using \code{mprotect} and runs on any processor architecture running Linux. Benchmark setup (2) installs all the necessary control data separation into the \code{.proconda} section but omits the not-yet existing hardware instruction to announce a legitimate write. It is basically identical to benchmark (1) but without calls to \code{mprotect} as software emulation.

The results of the software simulated runs are shown in~\autoref{fig:softref}. In these benchmarks, the runtime is largely dominated by the cost of syscall-induced context switches. Furthermore, we observe that the effect heavily varies between programs; upon manual inspection, we observed that the runtime increase is caused by frequent invocation of small functions, of which each invocation requires at least two \code{mprotect} calls to be simulated.  

To faithfully assess the effect of context switches on the overall runtime, we ran an additional set of benchmarks: during these runs, all the necessary control data separation and instrumentation steps were performed, but the WIO check instruction which announces a legitimate write was replaced by a \code{NOP} instruction to avoid the context switch. In essence, this set of benchmarks is identical to the previous one, but without the calls to \code{mprotect} that serve as a software emulation of \proconda's WIO checking phase. Hence, this set of benchmarks serves as a lower bound of performance impact of the approach provided sufficient hardware support is available. The results are of pictured in~\autoref{fig:hwref}.

\subsection{Micro-Benchmark}
Our software simulation requires the use of two \code{mprotect} calls per control-flow relevant data write, and therefore its performance impact becomes more apparent in cases with frequent function invocations (requiring that the \code{lr} register is stored). Consequently, the comparison of the software simulation and the reference run shows that the overhead of the syscall-based simulation becomes overwhelming in these cases. To better understand the absolute runtime impact of a software simulated WIO check we run a micro benchmark containing only WIO check invocations. The benchmarks consists of a 50000 sequences of \code{mprotect}-simulated WIO checks for which the absolute execution time is measured. On average, the the benchmarks took 6.45$\mu$s to complete on average, with a minimum runtime of 5.93$\mu$s and a maximum runtime of 7.72$\mu$s.   

\subsection{Memory Overhead}
\proconda introduces new instructions and a separate, specifically protected memory region into processes. This causes the code to become more bloated and additionally affects the memory requirements of affected programs. We analyzed both the increase of instructions due to WIO checks as well as the ELF section and program headers in memory during execution.

\begin{figure*}
	\subfloat[Increases in instructions count of \proconda-instrumented binaries.%
	\label{fig:meminst}]{%
		\includegraphics[width=0.45\textwidth]{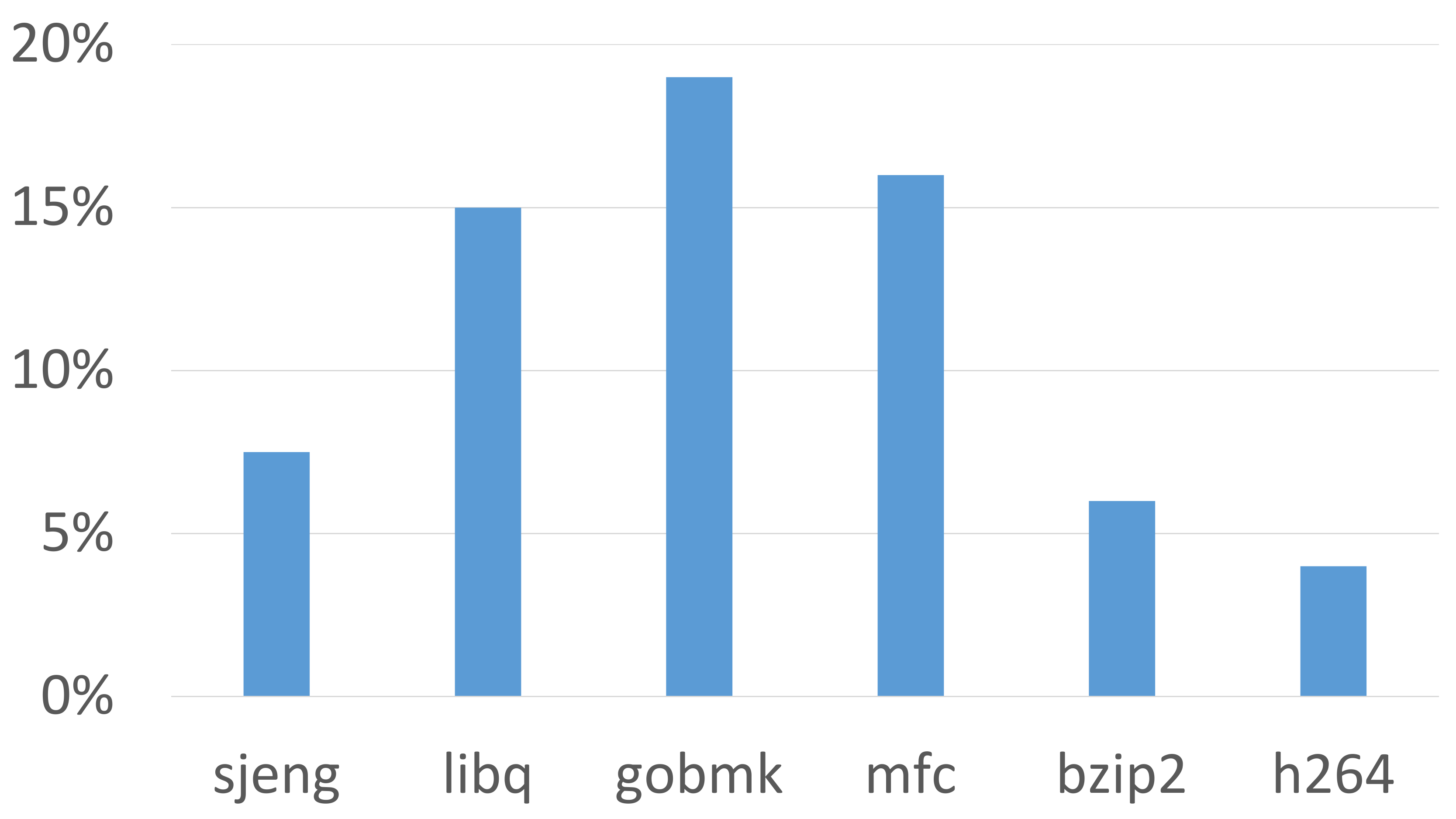}}
	\hfill
	\subfloat[Memory footprint of \proconda ELF files.%
	\label{fig:memelf}]{%
		\includegraphics[width=0.45\textwidth]{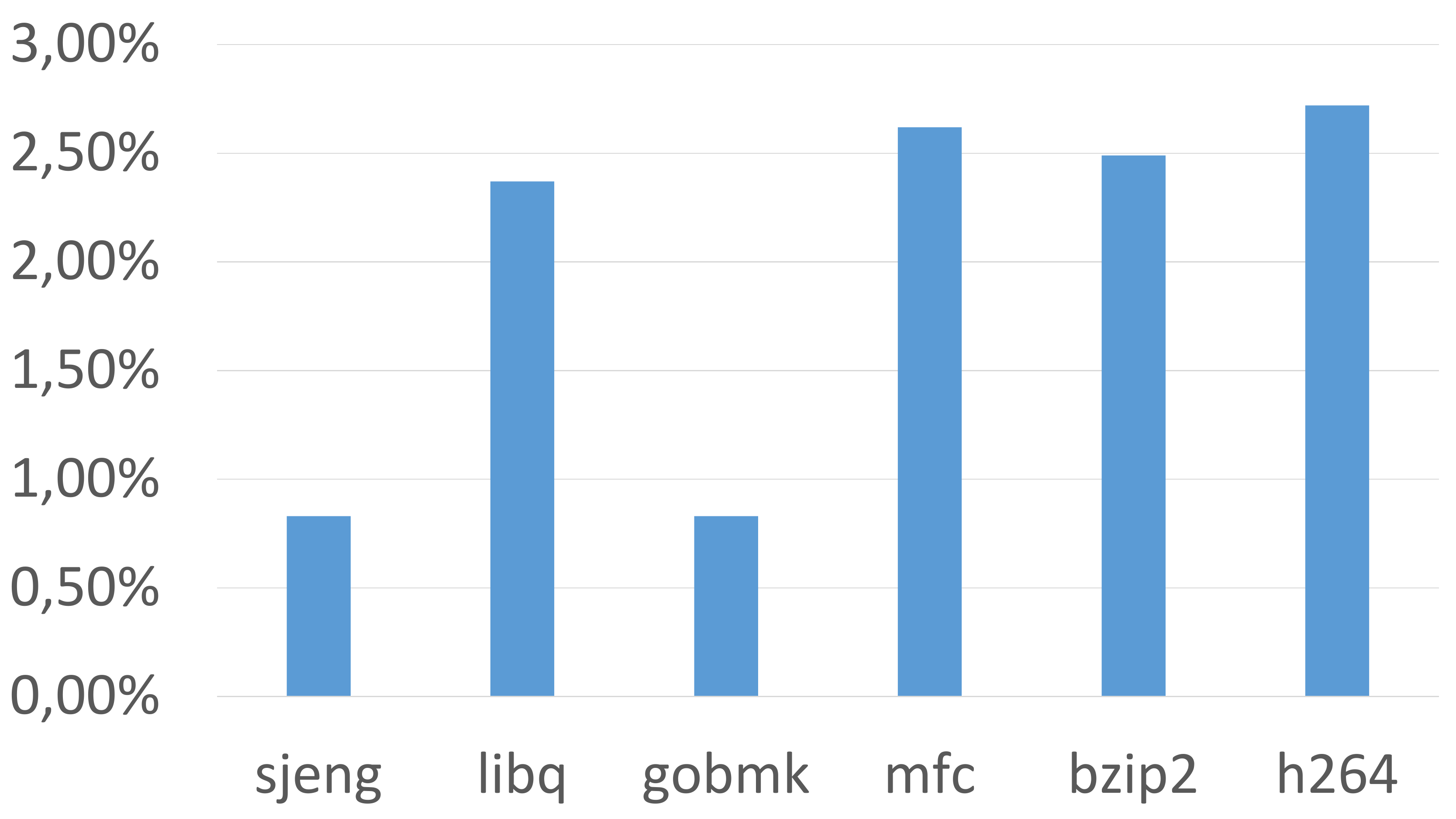}}
	\caption{\proconda memory overhead measured on SPEC CPU2006.}
\end{figure*}

Additional instructions where inserted during program load to initialize the protected memory segments, as well as during the WIO checks during runtime. \autoref{fig:meminst} illustrates the increases in instruction counts of the reference implementation to the unmodified binaries. As can be seen, the number of WIO checks varies heavily between 3.51\% on the lower and 19.66\% on the high end of the scale. On manual inspection, we found that heavily recurrent program structures with small functions such as in \texttt{gobmk} and \texttt{sjeng} are the most heavily affected. 

The increases of the memory footprint produced by the instrumented SPEC CPU2006 benchmarks is shown in~\autoref{fig:memelf}. This value incorporates both the protected memory region (at least on page, i.e. 4096KB) and the additional instructions needed to handle the WIO checks. On average, a reference implementation's memory footprint was 1.96\% larger compared to a stock binary.

\subsection{Effectiveness in preventing unintended control data corruption}
\label{sec:evaluation:sec}

\begin{table*}
\centering
\begin{tabular}{| c | c | c |}
\hline
Application 	& Memory corruption vulnerability  & Prevented \\ 
\hline
ncompress 4.2.4 & Global buffer overflow & $\checkmark$  \\
polymorph 0.40 & Global buffer overflow  & $\checkmark$  \\
htget 0.93 & Local buffer overflow & $\checkmark$ \\
\hline
\end{tabular}
\vspace*{.3cm}
\caption{Exploit benchmarks with their respective vulnerabilities.}
\label{tbl:vulns}
\end{table*}
\proconda protects control data by allowing only intentional writes to modify such data. Writes that were never supposed to touch control data are detected and their access is prevented before any data is tampered with.
To evaluate the effectiveness of the \proconda protection mechanism, we first evaluate the amount of code coverage achieved by \proconda's analysis before evaluating its ability to protect against memory corruption exploits.

The reason for the coverage evaluation is that like all dynamic approaches, \proconda requires that the test inputs for a program, i.e. the test cases, achieve a high code coverage, i.e. that all relevant, executable program paths containing control data write instructions are actually executed. The quality of the test cases therefore heavily impacts the effectiveness of the approach in real-world use cases. The test cases used by the \proconda analysis, including the Spec CPU2006 benchmarks, all achieved over 92\% basic block coverage~\cite{gove07:evaluating}.

To evaluate the effectiveness of \proconda in preventing illegitimate writes to control data, we applied \proconda to programs with well-documented security violations (cf.~\autoref{tbl:vulns}). The vulnerabilities in these samples represent a mixture of one or more write overflows on local and global control flow values. \proconda successfully prevents the corruption of control data in all cases.

%% file: sections/relatedwork.tex
\section{Related Work}
\label{sec:relatedwork}
Over the last few years, code reuse attacks and their mitigation has resembled an ongoing cat and mouse game.
Many mitigation techniques address the code reuse problem by trying to enforce the intended control flow of  a program (\cite{abbuer05:cfi,akcara08:WIT,zhse13:cfiforcots,zhwezh13:practicalcfi,tiroco14:forwardedgecfi,jatale14:safedispatch,dahapa15:hafix,mabibo15:ccfi,mabibo15:ccfi,molabr15:opaquecfi,chchat16:HCFI}) or by protecting the return pointers of the program against manipulation (\cite{cobejo03:pointguard,kuszpa14:cpi}).

Based on the ground-breaking work by Abadi et al.~\cite{abbuer05:cfi}, a variety of Control-Flow Integrity (CFI) approaches has been proposed that improve different aspects of CFI enforcement.
BinCFI~\cite{zhse13:cfiforcots} and CCFIR~\cite{zhwezh13:practicalcfi} are practical, low-overhead approaches that aim at protecting binaries by enforcing a coarse-grained form of CFI. Unfortunately, due to their coarse-grained nature, both techniques can be bypassed as shown by G\"oktas et al.~\cite{goatbo14:outofcontrol} and Davi~et~al.~\cite{dasale14:stitchingthegadgets}.
VTV~\cite{tiroco14:forwardedgecfi} and SafeDispatch~\cite{jatale14:safedispatch} are two fine-grained, compiler-level CFI approaches that aim specifically at protecting vtables. While both are able to reliably protect against COOP attacks~\cite{scteli15:coop}, they provide no security guarantees against classical ROP exploits~\cite{dasale14:stitching}.
HAFIX~\cite{dahapa15:hafix}, Opaque CFI (O-CFI)~\cite{molabr15:opaquecfi} and PointGuard~\cite{cobejo03:pointguard} are approaches that rely on information hiding mechanisms to efficiently enforce control flow. 
As stated by Abadi et al.~\cite{abbuer05:cfi} and later on underlined by Evans et al.~\cite{evfigo15:missingthepointer}, information hiding schemes trade strong security guarantees for a lower runtime overhead, but are therefore prone to getting circumvented eventually.

In contrast to the aforementioned schemes, which focus on keeping the contents of control-flow relevant elements safe by hiding them from an attacker, Mashtizadeh et al.~\cite{mabibo15:ccfi} propose Cryptographic CFI (CCFI), an approach that protects the integrity of control-flow relevant elements using message authentication codes (MACs). 

One of the main problems in the practical deployment of CFI is the trade-off between overhead and security, i.e. coarse-grained approaches work faster and are therefore more practical, but also more insecure, whereas fine-grained approaches usually incur such a high overhead that their practical deployment becomes impossible.
Therefore, Kutznetsov et al.~\cite{kuszpa14:cpi} have introduced \textit{Code Pointer Integrity (CPI)}, an approach that aims to enforce memory safety for all code pointers in a program while keeping a low overhead. 

\proconda follows the same idea as \textit{CPI} concerning the separation of code pointers into a separate memory segment. However, the mechanism by which CPI secures these pointers is based on information hiding. Recent research by G\"okta\c{s}  et al.~\cite{goektas16:undermining} and Oikonomopoulos et al.~\cite{oikonomopoulos16:poking} independently show serious, generic attacks on entropy-based information hiding schemes, concluding that the trust in defenses based on hidden memory regions is misplaced.
In contrast to the CPI approach, \proconda does not rely on information hiding to secure memory segments, thus circumventing the vulnerabilities of information hiding and data pointer overwrites as described Evans et al.~\cite{evfigo15:missingthepointer}. Instead, we create a map between code sections and pointers and implement a form of  access control that only allows legitimate code sections to change the respective pointers.

In this regard, \proconda is most similar to the \textit{Write Integrity Testing (WIT)} approach by Akritidis et al.~\cite{akcara08:WIT}. WIT prevents instructions from modifying objects they are not allowed to modify by first computing the control-flow graph and the set of objects that can be written by each instruction at compile time. Then it assigns a color to each write instruction and each object in the program and uses the resulting color table in an additional compilation phase to insert checks into the code to compare the color of each write instruction with the color of the memory location it tries to write to. If the colors do not match, a security exception is raised. 
\proconda also enforces write integrity by matching instructions with the set of objects they are allowed to write to. In contrast to WIT, we do not use a coloring scheme to check for possible violations but instead employ access control mechanisms. 

\proconda is the to the best of our knowledge the first approach to combine \textit{CPI} principles with hardware supported memory access control and write instruction origin checking.

%% file: sections/conclusion.tex
\section{Conclusion}
\label{sec:conclusion}
This paper presents \proconda, a novel technique to defend against the root cause of memory corruption exploits: illegitimate writes that manipulate control flow. \proconda effectively protects control data by write origin checking without relying on information hiding.
We implemented a software-based prototype of the \proconda enforcement mechanism on an ARM CPU running a Linux kernel. This prototype statically analyzes the assembler files of a program to identify memory addresses and registers containing control-flow relevant data and uses this information to detect write attempts to this data using backwards slicing. All legitimate write accesses to control-flow relevant data are then encapsulated with \code{mprotect} syscalls that regulate access to a special, write-protected memory region, in which all control-flow relevant elements are stored. %
The software-based emulation of the \proconda enforcement is effective on real-world exploits and incurred an runtime overhead ranging from 40\% to 300\% and an average memory overhead of 2\% on the popular SPEC benchmark. A separate evaluation using a micro benchmark has shown that this overhead can be almost wholly attributed to our use of \code{mprotect} syscalls to simulate a hardware-supported write access control mechanism. A reference implementation simulating the usage of proper hardware support incurred, on average, a runtime overhead of 4\%.

%% file: sections/controldataidentification.tex
\section{Identifying control data memory locations}
\label{appendix:cdi}
While the example given in \autoref{subsec:design1} illustrates the case where a control-flow destination stems from memory directly, there are also cases where there is
an arbitrary number of registers in between to carry a value from memory to a control-flow relevant instruction. 
For instance, \code{BX r5} (`Extended Branch to r5') jumps to the address stored in the register \code{r5}. At an earlier point in the program, the value of \code{r5} must either be loaded from memory (.e.g, \code{LDR r5,=0x0a130000}) or calculated as an offset to the program counter \code{pc}. 
In case the register is loaded from memory, the corresponding memory address must then contain control data. 
Another, although rarer case, is the use of arithmetics to calculate the code address. 
In compiled ARM code, this can only be seen in optimized jump tables. For example, \code{ADD pc, r3, \#8} can be used to represent a \code{switch(r3)} C-statement in which each possible \code{case}-target for r3 is 8 bytes apart. 
In this case, \code{r3} can be considered control data itself as it can influence \code{pc} to jump arbitrarily in multiples of 8.

\proconda handles these cases by tracking each register used in a control flow instruction back to a memory location (in the form of an offset to the stack base), which is then marked as containing control data. \autoref{tab:cfinstructions} gives an overview of all ARM instructions that change control flow. Note that on ARM, it is indeed possible to perform arithmetic operations on control-flow relevant registers like the program counter, \emph{pc}. This makes ARM an interesting target architecture for approaches like \proconda that focus on protecting the integrity of a program's control flow.

\begin{table}[hbtp]
	\centering
	\begin{tabular}{ l | l }
		\textbf{Instruction} & \textbf{Meaning} \\
		\hline
		\code{B, BX} & Branch \\
		\code{BL, BLX} & Branch and Link \\
		\code{POP \{pc\}} & Branch and Link \\
		\code{LDR pc\, [...]} & Branch and Link \\
		\code{ADD pc, r1, r2} & Jump to \code{r1} $+$ \code{r2} \\
		\code{ADD pc, pc, r2} & Relative jump by \code{r2} bytes
	\end{tabular}
	\vspace*{0.3cm}
	\caption{ARM Control Flow Instructions that operate on registers or RAM.}
	\label{tab:cfinstructions}
\end{table}

%% file: sections/armassembly.tex
\section{Targeting the ARM Assembly}
\label{appedinx:taa}
In the following, we will first explain the advantages of choosing assembly as a program representation for our analysis before elucidating our decision to use ARM as the underlying hardware architecture for our prototype.

\subsection{Targeting Low-level Assembly}
In order to keep our prototype independent of the specifics of a programming language or compiler, our static analysis phase works directly on the assembler files of the program. Since the structure and the instructions in assembler code only depend on the underlying architecture and not on the programming language that was used to create the original program, our \proconda prototype can handle programs written in a variety of languages. Moreover, we avoid the problem of programs implementing undefined behavior, which is then handled differently from compiler to compiler. 

Another compiler-related reason for choosing assembly as the underlying program representation for our analysis is that some control relevant data is only generated at the very end of a compilation pass (e.g. GOT pointers and jump tables). Consequently, these pointers are not part of the IR during the compilation, and would require additional modifications of the code after the code generation. Implementing our approach as part of a compiler suite would therefore severely affect the compilers architecture beyond their intended design and functionality. Targeting the program assembly, instead, does not affect the previous compilation steps and naturally fits into the usual sequence of compilation and linking as an immediate step. This is a trade-off: we can leave the compiler suite infrastructure intact, but instead of working on a very expressive and generic IR we have to tailor the changes \proconda introduces to specific architectures.

Last but not least, exploits are architecture-specific and manifest themselves differently based on the architecture the program was compiled for. Hence, we can use assembler code as a common target for a specific architecture. 

\subsection{Targeting the ARM Architecture}
Choosing ARM over x86 as the underlying architecture has a practical as well as a research-specific advantage.
From a practical point of view, ARM offers a fixed-length instruction set that facilitates the computation of offsets between different points in a program's assembler code. These offsets are used to identify the code locations of legitimate write instructions (represented as offsets from the main entry point or the last called function, respectively) and to instrument these instructions with code enabling the upcoming legitimate writes to the protected memory area.

From a research-specific point of view, ARM is the more interesting architecture for analyzing flows between control data as it allows arithmetic operations on control-flow relevant registers like 
\code{pc}. Thus, there are more interesting scenarios to run \proconda's analysis and enforcement mechanism against.

Despite the current focus on ARM, an adaptation of the prototype to x86 would only require minor changes regarding the instruction handling and only a small extension to our analysis. Lacking hardware support, the basic protection mechanism makes use of \code{mprotect} syscalls to allow legitimate instructions to write to the otherwise protected \proconda section. It is therefore independent of the underlying hardware architecture.

%% file: sections/programcoverage.tex
\section{Achieving Sufficent Program Coverage}
\label{appendix:aopc}

Ideally, what we would like to achieve in \proconda's \phasetwo~phase is known as full path coverage in the area of software testing, i.e. every path through a program should be executed at least once. However, path coverage and similar test coverage criteria have been shown to be infeasible in practice due to a number of reasons such as the presence of loops or dead code~\cite{zuhama97:testcoverage}. 

As finding and defining feasible coverage criteria along with the methods to fulfil them constitutes its own research area and is therefore out of scope for this paper, we instead use a simpler, more practical approach. 

To achieve the best possible coverage of all program paths that can actually be executed we select program inputs as follows:
For each input, we first determine the domain of the input, e.g. for an integer type input, we have possible inputs ranging from \code{INTEGER\_MIN\_VALUE} to \code{INTEGER\_MAX\_VALUE}. Based on the domain, we then select input values that not only are part of the domain itself but that also include boundary values, i.e. in the case of integers, we include both \code{INTEGER\_MIN\_VALUE} and \code{INTEGER\_MAX\_VALUE}. Afterwards, we combine these inputs into test sets for each program so that the best possible coverage is achieved given the restrictions imposed by the dynamic nature of our approach.

Since our inputs are regular and do not attempt to exploit the program in any way, the resulting accesses to the control-flow relevant addresses identified before and the instructions that execute them are considered legitimate.